\documentclass[aps, prb, twocolumn, superscriptaddress, reprint]{revtex4-2}
\pdfoutput=1

\usepackage[final]{graphicx}

\usepackage{textcomp}

\usepackage{color, colortbl}
\usepackage{microtype}
\usepackage[symbol,bottom]{footmisc}
\usepackage{subfigure}
\usepackage[colorlinks = true, allcolors = blue]{hyperref}

\begin{document}
\setlength{\parindent}{10pt}
\setlength{\textfloatsep}{20pt}

\title{Effects of the rf current and bias field direction on the transition from linear to non-linear gyrotropic dynamics in magnetic vortex structures} 

\author{Lakshmi Ramasubramanian}
\thanks{L. R. and V. I. contributed equally to this work}
\affiliation{Institute of Ion Beam Physics and Materials Research, Helmholtz-Zentrum Dresden-Rossendorf, 01328 Dresden, Germany}
\affiliation{Institute of Physics, Chemnitz University of Technology, 09107 Chemnitz, Germany}

\author{Vadym Iurchuk\footnotemark[1]}
\email{v.iurchuk@hzdr.de}
\affiliation{Institute of Ion Beam Physics and Materials Research, Helmholtz-Zentrum Dresden-Rossendorf, 01328 Dresden, Germany}

\author{Serhii Sorokin}
\affiliation{Institute of Ion Beam Physics and Materials Research, Helmholtz-Zentrum Dresden-Rossendorf, 01328 Dresden, Germany}

\author{Olav Hellwig}
\affiliation{Institute of Ion Beam Physics and Materials Research, Helmholtz-Zentrum Dresden-Rossendorf, 01328 Dresden, Germany}
\affiliation{Institute of Physics, Chemnitz University of Technology, 09107 Chemnitz, Germany}

\author{Alina Maria Deac}
\affiliation{Dresden High Magnetic Field Laboratory (HLD-EMFL), Helmholtz-Zentrum Dresden-Rossendorf, 01328 Dresden, Germany}

\date{\today}

\begin{abstract}
    We present a frequency-domain study of the dynamic behavior of a magnetic vortex core within a single Permalloy disk by means of electrical detection and micromagnetic simulations. When exciting the vortex core dynamics in a non-linear regime, the lineshape of the rectified dc signal reveals a resonance peak splitting which depends on the excitation amplitude. Using micromagnetic simulations, we show that at high excitation power the peak splitting originates from the nanosecond time scale quasi-periodic switching of the vortex core polarity. Using lock-in detection, the rectified voltage is integrated over a ms time scale, so that the net signal detected between the two resonant peaks for a given range of parameters cancels out. The results are in agreement with the reported effects of the in-plane static field magnitude on the gyration dynamics, and complement them by detailed analysis of the effects of the rf current amplitude and the azimuthal angle of the in-plane bias magnetic field. Systematic characterization shows that a transition from linear to nonlinear dynamical regime can be controlled by rf current as well as by varying the magnitude and the direction of the bias magnetic field. 
\end{abstract}


\maketitle 

\renewcommand\thesection{\arabic{section}}
\section{\label{title1}Introduction}
Magnetic vortices, which are spontaneously formed in magnetic disks of specific aspect ratios, are fundamental, topologically protected magnetic structures with an in-plane circulating magnetization and an out-of-plane magnetized vortex core (VC), which at remanence is localized in the center of the disk~\cite{aharoni_upper_1990,usov_magnetization_1993}. Since the first experimental observation of vortex formation in Py disks~\cite{shinjo_magnetic_2000}, the physics of magnetic vortices has evolved into a large and intense field of research that holds promises for future technological applications. Indeed, magnetic disks in the vortex state have the potential for application in biomedicine as magnetic microparticles that help to destroy cancer cells~\cite{kim_biofunctionalized_2010, leulmi_triggering_2015} or act as drug delivery agents~\cite{kim_mechanoresponsive_2011}. Asymmetric systems of exchange-coupled vortices can also serve as sources for the generation of short wave-length dipole-exchange spin-waves of a directional nature, with potential applications in magnonic computing~\cite{sluka_emission_2019}. Recently, rich non-linear spin-wave dynamics in a magnetic vortex was revealed including excitation of whispering gallery magnons via non-linear 3-magnon scattering processes~\cite{schultheiss_excitation_2019, koerber_nonlocal_2020} and VC gyration-assisted excitation of spin-wave frequency combs in the GHz range~\cite{heins_comb_2022} giving prospect for using magnetic vortices as magnonic reservoirs for unconventional data processing schemes.

In this paper, we present an extensive study of the non-linear gyration dynamics in a magnetic vortex investigated using homodyne electrical detection techniques. A similar experimental study of the non-linear gyration dynamics showed the in-plane bias field-induced resonance splitting \cite{kim_double_2013}. Based on a generic mechanical model of a nonlinear oscillator, the observed splitting was attributed to the dynamical bifurcation in the non-linear regime. Our work demonstrates that the origin of the resonant peak splitting is the nanosecond time-scale quasi-periodic VC switching at high excitation amplitudes, which can be achieved by varying not only the bias field magnitude, but also its azimutal angle as well. We show that the transition from the linear to the non-linear dynamical regime in single Py disk can be efficiently manipulated at constant rf excitation power by varying the external control parameters, i.e. the bias magnetic field magnitude and direction.

\begin{figure}[ht!]
\centering
\includegraphics[width=0.45\textwidth]{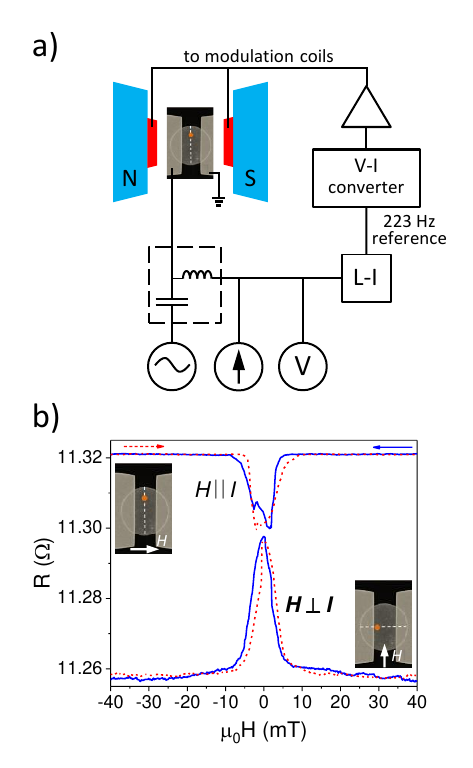}
    \caption{(a) Schematics of the measurement setup with magnetic field modulation of the magnetoresistance enabling rectified dc voltage measurements via lock-in technique. To generate a modulation field (collinear to the static bias magnetic field) at the lock-in reference frequency (here 223 Hz), copper wire coils (red) were attached to the electromagnet poles (blue). (b) Typical  magnetoresistance loops measured from a disk~(here radius $R=$ 4.4~µm, thickness $t=$ 35~nm)~for a dc current $I_{dc}=$ 5~mA). Blue solid (red dashed) lines are the backward (forward) field sweep measurements data from negative (positive) saturation as indicated by the corresponding arrow. The insets depict the schematic direction of the VC (orange dot) displacement depending on the orientation of the external field $H_b$.}
    \label{fig1}
\end{figure}

\section{\label{title2}Experimental methods}

Magnetic microdisks were fabricated on Si/SiO$_2$ substrates using electron beam lithography, followed by the deposition of a 35 nm Permalloy (Py = Ni$_{80}$Fe$_{20}$) film by e-beam evaporation and conventional lift-off. As the next step, Cr/Au contact pads were deposited to provide electrical access to each microdisk. We used a standard magnetotransport setup with rf capability (see Fig.~\ref{fig1}(a)) for electrical detection of magnetization dynamics in single magnetic vortices at room temperature. The detection technique exploits the anisotropic magnetoresistance (AMR) effect, i.e. the resistance change with the relative angle between the direction of the electrical current and the magnetic moment of a magnetic structure~\cite{stoehr}. An rf current injected through a bias-T into the microdisk device induces the gyrotropic motions of the VC at resonance, via the joint action of the spin-transfer torque and rf Oersted field, and thereby leads to an oscillating AMR provided the VC is displaced from the center of the disk. It was shown previously that in such devices, the VC gyration is primarily driven by the Oe field with a minor contribution from the spin transfer torque~\cite{kim_double_2013}. The time-averaged product of the rf current and the dynamical resistance – which results in a rectified dc voltage – is measured by a conventional homodyne detection scheme using a lock-in amplifier. To improve the signal-to-noise ratio, magnetic field modulation at the lock-in reference frequency was used~\cite{goncalves_spin_2013}. The details on the sample fabrication process, as well as the experimental procedure for electrical detection, have been reported elsewhere~\cite{paper}. 
One has to note, that we use Py disks with relatively large radii (3.2 µm and 4.4 µm). These relatively large sizes were initially chosen for Cr$^+$ ion implantation in order to build double output frequency devices. The results will not be covered here, as they have been reported elsewhere~\cite{paper}.

\begin{figure*}[ht!]
    \includegraphics[width=0.9\textwidth]{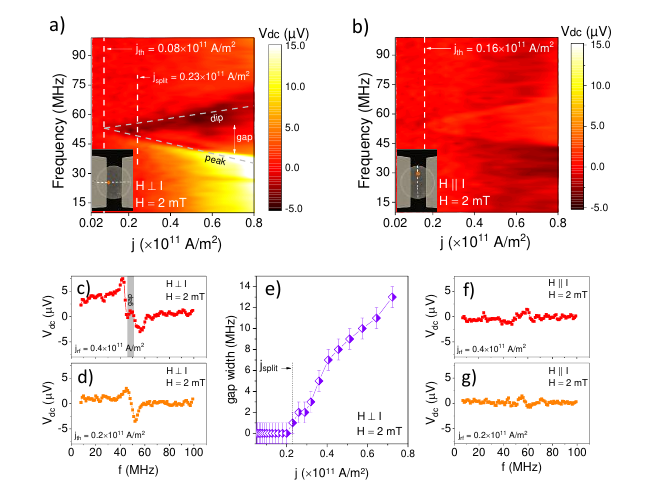}
    \caption{Spectral color maps of the excitation power-dependent rectified dc signal measured of a Py disk ($R=$ 3.2 $\mu$m, $t=$ 35 nm) taken at a static field $\mu_0 H_b$ = 2 mT for (a) $H_b \perp I$ and (b) $H_b \parallel I$. The insets show the schematic position of the VC for the given field orientation. Threshold current $j_{th}$ (marked by a vertical dashed line in (a) and (b)) corresponds to the current density for which the signal peak can be clearly detected. Dashed sloped lines in (a) are guides to the eye showing schematically the peak/dip frequency shifts with increasing $j_{rf}$. $j_{split}$ corresponds to the current density where the frequency splitting is detected. (c,d) Single $V_{dc}$ spectra measured at high/low rf current density ($j_{rf}=$ 0.4 $\cdot 10^{11}$ A/m$^2$ and $j_{rf} =$ 0.2 $\cdot 10^{11}$ A/m$^2$ respectively) for $H \perp I$. e) Zero-signal gap for $H \perp I$ as a function of $j_{rf}$. (f,g) Single $V_{dc}$ spectra measured at the same values of $j_{rf}$ as in (c,d) for $H \parallel I$.}
    \label{fig2}
\end{figure*}

\section{\label{title5}Results and discussion}

\renewcommand\thesubsection{\thesection.\arabic{subsection}}
\subsection{Detection of the vortex core nucleation using static magnetoresistance measurements}

We first confirmed the presence of a magnetic vortex in the fabricated disks by measuring the static magnetoresistance (MR). It has to be noted that in order to reveal a measurable MR signal, a relatively large dc current (about 50 mA) had to be applied through the as-prepared devices in order to break the thin Py oxide layer formed during the fabrication process, which isolates the magnetic disk from the contact pads. As the oxide does not break homogeneously under the contact pad, this leads to a presumably non-uniform electrical current distribution within the disk, whose effects on the dynamics are discussed further in the text. Fig.~\ref{fig1}(b) shows the MR of the disk at a dc current I$_{dc}=$ 5~mA measured for in-plane bias magnetic field $\mu_0 H_b$ applied parallel (top curve) and perpendicular (bottom curve) to the direction of the probing current $I$ .

Starting from positive in-plane saturation field, i.e. with the disk in a single domain state, the external field is reduced to nucleate a vortex in the disk. This is observed in our measurements as a resistance decrease (increase) for $H_b \parallel I$ ($H_b \perp I$). Further sweeping the field towards negative values displaces the VC within the disk (in the direction perpendicular to the applied field) until its expulsion, which is accompanied by a recovery of the resistance value of the saturated state. The same measurement is repeated in the reverse direction (from negative to positive saturation field) to reproduce the vortex nucleation/expulsion behaviour. 

\subsection{Vortex core dynamics at large excitation power: resonant bistability}

\renewcommand\thesubsubsection{\thesubsection.\arabic{subsubsection}}
\subsubsection{Vortex resonance splitting in strongly nonlinear regime}

The gyrotropic motion of the VC is probed by frequency-resolved homodyne detection using the spin rectification effect. 
The rf frequency sweeps were performed varying one of the three external parameters, i.e. the amplitude of the excitation rf current density $j_{rf}$, the bias magnetic field magnitude $H_b$ or the in-plane azimuthal angle $\phi_H$ of the external bias field while keeping the other two constant. Here, the value of the current density $j_{rf}$ injected into the device can only be estimated taking into account the impedance mismatch between the device and the 50$\Omega$ circuit. More specifically, we define $j_{rf}$ as $\frac{1}{w t} \sqrt{P_{rf} (1-\Gamma^2) / R_0}$, where $w$ is the width of the pad in contact with the disk, $t$ is the thickness of the disk, $R_0$ is the static resistance of the disk, $P_{rf}$ is the rf power delivered to the sample from the signal generator and $\Gamma = | \frac{R_0-50\Omega}{R_0+50\Omega} |$ is the reflection coefficient due to the impedance mismatch between the sample and the circuit. In this section, we focus on the effect of the $j_{rf}$ amplitude on the VC dynamics.

To understand the dynamic behaviour of the VC at high excitation power, we first performed frequency sweeps at a fixed in-plane bias field $\mu_0 H_b$ = 2 mT (for $H_b \perp I$) while increasing rf power (see Fig.~\ref{fig2}(a)) in order to observe detectable gyration dynamics signal. A first clear measurable response was observed at a threshold current $j_{th}=$ 0.08 $\cdot 10^{11}$ A/m$^2$. With increasing $j_{rf}$, the rectified signal also increases as can be seen from the intensity map of Fig.~\ref{fig2}(a), since the VC now gyrates with a larger radius, leading to a larger MR change. Fig.~\ref{fig2}(d) shows a clear antisymmetric lorentzian-like signal~\cite{goto_electrical_2011} acquired at $j_{rf} =$ 0.2 $\cdot 10^{11}$ A/m$^2$. Another striking feature appears when reaching $j_{split}=$ 0.23 $\cdot 10^{11}$ A/m$^2$, where the dc voltage signal evolves from a single antisymmetric peak response to a 'peak-dip' lineshape separated by a plateau (gap) between them (Fig~\ref{fig2}(a)). This zero-signal gap expands with increasing $j_{rf}$, making the splitting in the resonance lineshape more pronounced (see Fig.~\ref{fig2}(a,e)). The splitting behaviour is considered as an indication of the vortex entering a non-linear bistability region. This effectively means that in the strongly non-linear regime, the system becomes resonantly stable at two excitation frequencies.

The clear double-peak response is not reproduced for $H_b \parallel I$ (see Fig.~\ref{fig2}(b,f,g)). Although in Fig.~\ref{fig2}(b) one can visually outline a similar triangular-shaped region of the constant rectified voltage at high current densities, the detailed analysis shows that the $V_{dc}$ spectra in this region are qualitatively different for the $H_b \perp I$ case. Indeed, for this field orientation, the threshold current for the dc voltage detection is 0.18 $\cdot 10^{11}$ A/m $^2$, which is 2 times higher as compared to the $H_b \perp I$ case. Moreover, for $H_b \parallel I$, the rectified signal amplitudes are almost one order of magnitude lower, and the peak splitting at high $j_{rf}$ is not clearly developed (see Fig.~\ref{fig2}(f)).

\begin{figure}[b]
    \includegraphics[width=0.5\textwidth]{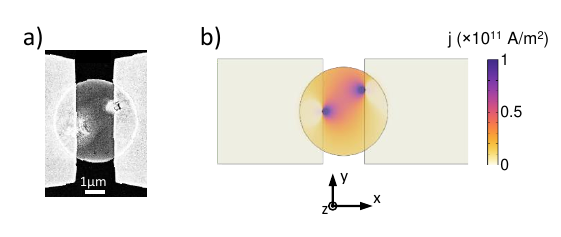}
    \caption{(a) SEM image of the shorted device after $\sim$60 mA dc current was passed through the disk. The brighter SEM contrast reveals the areas where the Permalloy oxide was broken and the current density within the disk was the highest. (b) Simulated distribution of the current density in a 35 nm thick Py disk ($R$ = 2.1 $\mu$m) calculated using COMSOL Multiphysics\textsuperscript{\textregistered}. The scale bar shows the current density in A/m$^2$. In this simulation, a 5 mA dc current is passed through the gold pads ohmically contacted to the Py disk. The experimentally observed breaking of the Permalloy oxide layer is mimicked by considering the current injected into the disk through the point-like regions at the edges of the gold pads (here taken randomly as circular areas with 20 nm radii). The result is the non-homogeneous distribution of the injected current where higher current density is observed closer to the contact pads.}
    \label{fig3}
\end{figure}

\begin{figure*}[ht!]
    \includegraphics[width=0.8\textwidth]{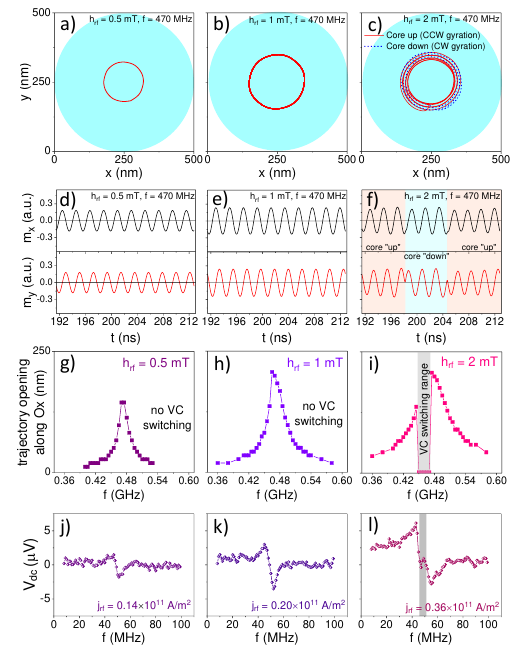}
    \caption{(a--c) Simulated trajectories of the VC for a Py disk ($R$ = 250 nm; $t$ = 30 nm) at $f$ = 470 MHz and $h^x_{rf}$ = 0.5, 1 and 2 mT. (d--f) Corresponding time traces of the $m_x$ (top) and $m_y$ (bottom) components of the normalized magnetization recorded for the last 10 periods ($\sim$21.3 ns) of the simulation time. (g--i) Trajectory opening vs. excitation frequency for $f$ = 470 MHz and $h^x_{rf}$ = 0.5, 1 and 2 mT. (j--l) Experimental dc voltage spectra measured at $\mu_0 H_b$ = 2 mT for $H_b \perp I$ taken from the spectral map of Fig.~\ref{fig3}(a) at given $j_{rf}$ values. Shaded region in (i) and (l) refers to the VC switching range.}
    \label{Fig4}
\end{figure*}

This spacial anisotropy in the VC dynamics is attributed to the inhomogeneous current distribution within the magnetic disk. Fig.~\ref{fig3}(a) shows an SEM image of a successfully contacted device, taken after a large dc current of 60 mA was passed through leading to the on-purpose device shortcut. It is worth mentioning that the experimental data shown in Figs.~\ref{fig1},~\ref{fig2},~\ref{Fig4},~\ref{Fig5} and~\ref{Fig6} are obtained on devices different from the one shown in Fig.~\ref{fig3}(a), which was used to purposely create a shortcut and thus visualize the typical current distribution within the disk. The brighter contrast region within the disk indicates the current flow path, confirming the point-like oxide breaking and the corresponding inhomogeneous current distribution leading to the different dynamics response depending on the bias field orientation. Fig.~\ref{fig3}(b) shows the simulated current distribution in the disk assuming that the thin oxide layer on top of the Py film is broken so that it forms a point-like electrical contact. This assumption agrees with the experimentally observed drop of the device resistance from $\sim$130 $\Omega$ to $\sim$12 $\Omega$ at 50~mA dc current applied to the as-prepared disk in order to break the oxide layer and establish a direct conductive contact between the disk and the pads.  In general, the local current density closer to the contact pads is higher compared to the disk center. Therefore, when the VC is displaced closer to the contact pads (i.e. when $H_b \perp I$), the local rf Oe field (generated by the injected rf current) driving its dynamics is expected to be higher as compared to the $H_b \parallel I$ case (when the VC is located far from the contract pads). This argument also explains the differences in the threshold current values and the peak splitting behavior in the experimental data of Fig.~\ref{fig2}.

A bistable behaviour similar to the one observed here in a single disk has previously been reported on an array of $\sim$2000 magnetic vortex structures probed by microwave reflection measurements at high excitation amplitudes ~\cite{buchanan_driven_2007}. Electrical detection of the double resonance response of the VC dynamics has also been reported, where such drift from linear gyrotropic motion is generally linked with non-linear phenomena that can be understood by considering the system as a non-linear oscillator for which two stable dynamical states co-exist (one with large and one with small oscillation amplitude)~\cite{kim_double_2013}. However, our analysis shows that in single magnetic disks, the double-resonance behavior and the appearance of a zero-signal shoulder between the two resonances originates from quasi-periodic switchings of the VC under extremely strong rf current excitation in agreement with the non-linear analytical model of the vortex dynamics~\cite{guslienko_nonlinear_2010}. This mechanism is revealed using micromagnetic simulations and is explained in detail in the next section.

\subsubsection{Dynamical micromagnetic simulations}

We use the GPU-accelerated MuMax3 software package~\cite{vansteenkiste_design_2014} for simulating the magnetization dynamics in the Py magnetic disks. We study the trajectories of the vortex gyrotropic mode as a function of the frequency of the rf magnetic field excitation. Similar to~\cite{iurchuk_stress_2021}, the VC is excited by an in-plane sinusoidal rf magnetic field $h_{rf} = h \sin{(2 \pi ft)}$, where $f$ is the excitation frequency, and $h=(h^x_{rf},0,0)$. The magnetization dynamics in the magnetic disk is simulated for different rf field amplitudes $h^x_{rf}$ over 100 gyration periods $T_0 = \frac{1}{f}$. The VC trajectory over the last 10 periods is recorded for each value of the excitation frequency $f$ in the chosen range. We use the following Py material parameters for the magnetization dynamics simulations: saturation magnetization $M_s$ = 800 kA/m, exchange constant $A_{ex}$ = 13 pJ/m$^3$, and damping parameter $\alpha$ = 0.008~\cite{iurchuk_stress_2021}. To reduce the simulation time, we consider a much smaller Py magnetic disk of 250 nm radius and 30 nm thickness discretized into 128×128×1 cells. Although the corresponding VC gyration resonance frequency for such disk ($\sim$ 470 MHz) is approximately one order of magnitude higher compared to the experimentally studied Py disk ($\sim$ 49 MHz), the evolution of the VC resonance behavior as a function of the rf field amplitude is qualitatively similar.

Fig.~\ref{Fig4}(a,b) shows the simulated trajectories of the VC at $f$ = 470 MHz and $h^x_{rf}$ = 0.5 mT (a) and 1 mT (b). For 0.5 and 1 mT amplitudes the steady state gyration sets after $\sim$20 gyration periods with quasi-circular VC trajectories. The time traces of the $m_x$ and $m_y$ magnetization components taken after the steady state gyration sets in (see Fig.~\ref{Fig4}(d,e)) show harmonic oscillations over time. The maximum VC displacement (trajectory opening) in the $x$ direction is extracted from the simulations and is plotted versus the frequency in Fig.~\ref{Fig4}(g,h) for $h_x$ = 0.5 and 1 mT respectively. These VC displacement traces define the dynamical MR variation whose derivative is proportional to the rectified dc voltage signal detected during experiments at a given frequency (see Fig.~\ref{Fig4}(j,k)). Increasing the rf field amplitude from 0.5 mT to 1 mT, the resonance peak broadens and the trajectory opening (rectified signal amplitude) increases. However, for $h^x_{rf}$ = 2 mT, no steady state trajectory was observed, but instead we observe quasi-periodic switching of the VC polarity. For example, as seen in Fig.~\ref{Fig4}(c), the VC switches 2 times over the 10 gyration periods ($\sim$ 21.3 ns), from VC polarity "up" to "down" and back to "up", which is accompanied by a change in the gyration sense from counterclockwise to clockwise and back to counterclockwise. It can be also seen from the corresponding time traces of the $m_x$ and $m_y$ components (see Fig.~\ref{Fig4}(f)) that the phase of the oscillations shifts twice over the 10 gyration periods indicating the core switching events. Fig.~\ref{Fig4}(i) shows the frequency range for $h^x_{rf}$ = 2 mT in the vicinity of the gyration resonance where the VC undergoes multiple quasi-periodic polarity switchings. We find that these switchings occur via a known mechanism of vortex-antivortex pair formation~\cite{arekapudi_direct_2021} and the subsequent annihilation of the original vortex with the created antivortex, when the vortex core reaches its critical velocity ($\sim$333 m/s for Py~\cite{lee_universal_2008}). In our simulations, for $h^x_{rf}$ = 2 mT, (Fig.~\ref{Fig4}(c,f)) we observe quasi-periodic polarity switching events approximately every 3 gyration periods, in agreement with the previously reported results for Py disks~\cite{Gliga_2011}.

Core switching is accompanied by a change in the sense of gyration. Since the electrical detection method is sensitive to the VC polarity~\cite{kim_current-induced_2010, goto_electrical_2011, sushruth_electrical_2016}, when the vortex core flips millions of times over one lock-in time constant (250 ms), the dynamical resistance response will be counter balanced for each subsequent polarity switching event. This leads to a zero net rectification voltage for the rf frequencies when switchings take place and makes the antisymmetric resonance lineshape to appear like a peak followed by a dip with a pronounced flat plateau as seen in the corresponding experimental spectrum (in Fig.~\ref{Fig4}(l)). By increasing the excitation amplitude, the observed plateau widens significantly as now even off-resonant excitation is sufficiently strong to induce the VC polarity switching (see Fig.~\ref{fig2}(a)).

One has to comment, that in the frequency sweep experiments shown in this work, the equilibrium position of the vortex core is displaced with a static external field and an rf current is applied to the disk to excite the VC gyration. However, in the simulations, the vortex core is excited by an rf field and the gyration trajectories are recorded in the absence of any external bias field. Nevertheless, the experiments and simulations show good agreement as the vortex core is excited resonantly in both cases.

\subsection{Controlling the transition from linear to nonlinear dynamics at fixed rf current}

As seen from the above described measurements and simulations, the transition between linear and nonlinear dynamical regimes can be controlled by moving the VC between the regions in the disk with lower/higher rf current density (i.e. further/closer to the electrical contact pads) while keeping the input power constant. This can be realized by changing the magnitude of $H_b$ and/or the azimuthal angle $\phi_H$ of the external bias field.

\begin{figure}[hb]
    \includegraphics[width=0.5\textwidth]{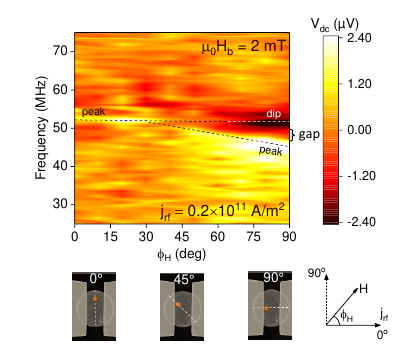}
    \caption{Spectral color map of the VC gyrotropic frequency vs. azimuthal angle $\phi_H$ measured for $j_{rf}$ = 0.2 $\cdot 10^{11}$ A/m$^2$ and $\mu_0 H_b$ = 2 mT. Dashed lines are guides to the eye. Insets at the bottom depict the schematical positions of the VC for $\phi_H$ = 0$^{\circ}$ (corresponding to $H \parallel I$), 45$^{\circ}$ and 90$^{\circ}$ (corresponding to $H \perp I$).}    
    \label{Fig5}
\end{figure} 

\subsubsection{$\phi_H$ dependence}

Fig.~\ref{Fig5} shows the spectral color map of the rectified dc voltage measured at fixed rf current $j_{rf}$ = 0.2 $\cdot 10^{11}$ A/m$^2$ and bias field $\mu_0 H_b$ = 2 mT for different azimuthal angles $\phi_H$ ranging from 0$^{\circ}$ ($H_b \parallel I$) to 90$^{\circ}$ ($H_b \perp I$). For $\phi_H$ = 0$^{\circ}$, a minuscule rectified signal is observed (note that this combination of $j_{rf}$, $H_b$ and $\phi_H$ corresponds to the spectrum shown in Fig.~\ref{fig2}(f)), as the VC is located between the contact pads (see bottom left inset for the schematic VC position) and the excitation power is slightly above the threshold rf current $j_{th}$ = 0.18 $\cdot 10^{11}$ A/m$^2$. When $\phi_H$ is increased to $\sim$15$^{\circ}$, a clear single peak is detected at about 52 MHz. It is attributed to the VC being displaced closer to the contact pads, where the VC is subjected to a stronger rf Oersted field excitation due to the inhomogeneous rf current distribution within the disk (see Fig.~\ref{fig3}). Note that here we refer to the Oe field generated by the current flowing along the \textit{z} direction, i.e. between the contact pads and the disk at the location of the point-like oxide breakthrough. For $\phi_H \geq$ 50$^{\circ}$, a pronounced antisymmetric signal is observed, whose amplitude increases with increasing the angle $\phi_H$, reaching its maximum at $\phi_H$ = 90$^{\circ}$. This behavior is consistent with the data of Fig.~\ref{fig2}(a) where the peak amplitude increases with increasing the rf current at $H_b \perp I$ (i.e. at fixed angle $\phi_H$ = 90$^{\circ}$), but here the rf current remains fixed, and the effective excitation is increased by increasing the azimuthal angle $\phi_H$ (i.e by moving the VC closer to the contact leads, see bottom middle and right insets).

One has to note that the resonance frequency slightly decreases with increasing angle, indicating a possible transition towards a non-linear gyration regime. Indeed, for angles close to 90$^{\circ}$, a small gap is observed between resonant peak and dip indicating a peak splitting behavior being a typical feature of the non-linear VC dynamics as was in detail explained in the previous section.
The presented data shows that the amplitude of the VC gyration (and even the transition form linear to non-linear dynamical regimes) can be controlled at constant excitation rf power and fixed bias magnetic field by changing the azimuthal field angle, i.e. by moving the VC between the regions of different effective excitation strength (here closer to or further away from the contact pads).

\subsubsection{$H_b$ dependence}

Up to now we discussed angular dependent and rf current dependent measurements carried out at a fixed bias magnetic field of 2 mT. We have shown that due to the inhomogeneous current distribution within the disk, the VC dynamical response is defined not only by the excitation power but by the spatial position of the VC with respect to the contact pads as well. It is expected that one can drive the VC in a non-linear regime at fixed rf current by shifting the VC towards the contact pads, thus increasing the effective Oe field excitation. To verify this, we have performed dynamical measurements at $H \perp I$ with $j_{rf}$ = 0.68 $\cdot 10^{11}$ A/m $^2$ for different values of bias field $H_b$. Here, we keep the azimuthal angle $\phi_H$ = 90$^{\circ}$ fixed 
(see the bottom right inset of Fig.~\ref{Fig5}). 

\begin{figure}[ht]
    \includegraphics[width=0.5\textwidth]{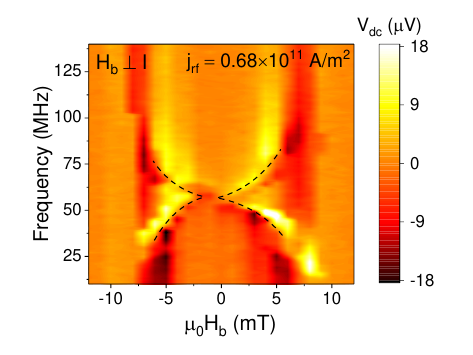}
    \caption{Spectral color map of the rectification signal $V_{dc}$ as a function of the VC gyrotropic frequency vs. external bias field $H_b$ measured for rf current density $j_{rf}$ = 0.68 $\cdot 10^{11}$ A/m$^2$ and $H_b \perp I$ (corresponding to $\phi_H$ = 90$^{\circ}$). Dashed lines are guides to the eye showing the increase of the "peak-dip" separation with increasing bias field.}
    \label{Fig6}
\end{figure}

Fig.~\ref{Fig6} shows the map of the rectified dc voltage frequency spectra acquired as a function of the bias field $H_b$ applied perpendicular to the rf current at $j_{rf}$ = 0.68 $\cdot 10^{11}$ A/m $^2$. At given current density and field azimuthal angle, a linear dynamical response is expected at low bias field (see Fig.~\ref{fig2}(a)), which is seen as a single asymmetric peak for $H_b \leq$ 1 mT. When $\mu_0 H_b$ is increased to $\sim$1.5 mT, an onset of the peak splitting is observed, which increases with increasing the bias field, leading to a pronounced "peak-dip" signal shape similar to the one shown in Fig.~\ref{fig2}(a,c) and Fig.~\ref{Fig5} for large $j_{rf}$ and $\phi_H$ respectively. This behavior indicates that increasing bias field drives the system to the non-linear dynamical regime, as the VC is displaced to the region with larger rf current density, i.e. with higher effective excitation power. At $\mu_0 H_b\sim$ 8 mT, the signal drops abruptly, corresponding to the VC expulsion from the probed region of the disk. When the field is swept to negative values, the behavior is reproduced almost symmetrically. The minor asymmetry with respect to the bias field may be attributed to a slightly off-centered position of the VC at zero bias field due to disk imperfections and/or residual magnetic fields between the electromagnet poles. For $H \parallel I$ (i.e. when the VC is located in the region with lower current density), no peak splitting was observed (the VC gyration remains within the linear regime) due to insufficient effective excitation power for the VC dynamics.

Similar to the $\phi_H$ dependence, this measurement shows that by moving the core between the regions of different effective excitation strength, one can control the dynamical regime of the VC gyration and switching from linear to non-linear dynamics at fixed rf excitation and bias field direction by varying only the magnitude of the bias magnetic field. This fixed magnetic field can also be introduced for instance by adding an adjacent magnetic layer coupled magnetostatically to the disk in a vortex state.

\section{Conclusions}

We have studied the VC nonlinear dynamics in a single magnetic microdisk as a function of external control parameters, i.e the excitation rf current, the static bias magnetic field magnitude and the in-plane field orientation. Via magnetotransport measurements, we have found that the effective excitation power depends on the VC position within the disk and is essentially higher for magnetic fields aligned perpendicular to the rf current flow. This anisotropy of the dynamical response is attributed to the inhomogeneous rf current distribution within the disk, leading to higher rf current-induced Oersted field for the case when the VC is located close to the contact pads. This effect enables tuning the transition from linear to non-linear dynamical regimes by moving the VC between the regions with effectively lower/higher excitation amplitude (by changing the bias field magnitude and/or in-plane angle) while keeping the nominal rf current flowing through the disk fixed. Furthermore, with higher resolution lithographic design of the contact pads, it appears to be possible to control the exact location of the oxide breakthrough point in order to optimize the parametric tunability of the transition from linear to non-linear vortex dynamics.

The described complex dynamics may be of use in multifrequency rf detectors based on magnetic vortices with two different working regimes (corresponding to the linear and non-linear dynamical regimes of the magnetic vortex). Depending on the selected dynamical regime, the magnetic vortex can be resonantly excited by either a single input frequency (linear regime) or a combination of two input frequencies (non-linear regime). In the non-linear case, the sensitivity to the difference between the input frequencies is defined by the plateau between the corresponding resonances and can be tuned by varying the external control parameters, i.e. excitation power and/or bias magnetic field magnitude/direction.

\begin{acknowledgments}
 This work is supported by the European Union's Horizon 2020 research and innovation programme under grant agreement No. 737038 (TRANSPIRE), the Helmholtz Young Investigator Initiative Grant No. VH-N6-1048, and the HLD at HZDR, member of the European Magnetic Field Laboratory (EMFL). Support by the Nanofabrication Facilities Rossendorf (NanoFaRo) at the IBC is gratefully acknowledged. The authors acknowledge fruitful discussions with Attila K\'akay.
\end{acknowledgments}



\bibliographystyle{apsrev4-2}
\bibliography{References.bib}

\end{document}